\documentclass[conference]{IEEEtran}
\IEEEoverridecommandlockouts
\usepackage{cite}
\usepackage{amsmath,amssymb,amsfonts}
\usepackage{algorithmic}
\usepackage{graphicx}
\usepackage{textcomp}
\usepackage{xcolor}
\usepackage{etoolbox}
\usepackage{booktabs}
\usepackage{multirow}
\usepackage[most]{tcolorbox}
\tcbuselibrary{skins,breakable}
\usepackage{url}

\AtBeginEnvironment{thebibliography}{%
  \setcounter{enumi}{0}%
  \setcounter{enumiv}{0}%
}

\pretocmd{\thebibliography}{\setcounter{enumi}{0}\setcounter{enumiv}{0}}{}{}
\makeatother

\def\BibTeX{{\rm B\kern-.05em{\sc i\kern-.025em b}\kern-.08em
    T\kern-.1667em\lower.7ex\hbox{E}\kern-.125emX}}

\makeatletter
\newcommand{\linebreakand}{
  \end{@IEEEauthorhalign}
  \hfill\mbox{}\par
  \mbox{}\hfill\begin{@IEEEauthorhalign}
}
\makeatother

\begin{document}

\title{Contextual Graph Embeddings: Accounting for Data Characteristics in Heterogeneous Data Integration\\
}

\author{\IEEEauthorblockN{Yuka Haruki}
\IEEEauthorblockA{\textit{Graduate School of Engineering} \\
\textit{The University of Tokyo}\\
Tokyo, Japan \\
haruki-yuka596@g.ecc.u-tokyo.ac.jp}
\and
\IEEEauthorblockN{Shigeru Ishikura}
\IEEEauthorblockA{\textit{Platform Business Marketing Department Data Operations Section} \\
\textit{Infomart Corporation}\\
Tokyo, Japan \\
s.ishikura@infomart.co.jp}
\linebreakand
\IEEEauthorblockN{Kazuya Demachi}
\IEEEauthorblockA{\textit{Platform Business Marketing Department Data Operations Section} \\
\textit{Infomart Corporation}\\
Tokyo, Japan \\
k.demachi@infomart.co.jp}
\and
\IEEEauthorblockN{Teruaki Hayashi}
\IEEEauthorblockA{\textit{Graduate School of Engineering} \\
\textit{The University of Tokyo}\\
Tokyo, Japan \\
hayashi@sys.t.u-tokyo.ac.jp}
}

\maketitle

\begin{abstract}
As organizations continue to access diverse datasets, the demand for effective data integration has increased. Key tasks in this process, such as schema matching and entity resolution, are essential but often require significant effort. Although previous studies have aimed to automate these tasks, the influence of dataset characteristics on the matching effectiveness has not been thoroughly examined, and combinations of different methods remain limited. This study introduces a contextual graph embedding technique that integrates structural details from tabular data and contextual elements such as column descriptions and external knowledge. Tests conducted on datasets with varying properties such as domain specificity, data size, missing rate, and overlap rate showed that our approach consistently surpassed existing graph-based methods, especially in difficult scenarios such those with a high proportion of numerical values or significant missing data. However, we identified specific failure cases, such as columns that were semantically similar but distinct, which remains a challenge for our method. The study highlights two main insights: (i) contextual embeddings enhance the matching reliability, and (ii) dataset characteristics significantly affect the integration outcomes. These contributions can advance the development of practical data integration systems that can support real-world enterprise applications.
\end{abstract}


\section{Introduction}
With the digitalization of industrial and economic systems, organizations are generating and accumulating vast and heterogeneous datasets \cite{KnowledgeDiscovery21,AlternativeData24}. Although such data have significant potential for creating new value through integration, their diverse origins often lead to inconsistencies in structure, format, and semantics. As a result, data integration has emerged as a critical enabler of inter-organizational collaboration, supporting coherent analysis and reliable decision-making \cite{Issues24,Inter-Organizational24}.

However, datasets originating from various sources often exhibit different formats and structures, which result in inconsistencies when they are integrated. Consequently, preprocessing steps that are designed to remove contradictions and redundancies and to create unified and consistent datasets have become crucial \cite{DataIntegration17}. Among data integration tasks, schema matching (SM), which refers to identifying correspondences between attributes across datasets, and entity resolution (ER), in which records that refer to the same real-world entities are consolidated, are the most fundamental \cite{OntologyMatching04,ReferenceReconciliation05}. However, these tasks are traditionally labor intensive, relying on the manual design of rules and expert domain knowledge \cite{Wordnet95}. Therefore, substantial research has focused on automating these tasks to reduce costs, ranging from rule-based heuristics and knowledge-based methods to the more recent machine learning \cite{Principles12,SurveyofBigDataManagement16} and graph embedding \cite{EmbDI20,IDAGEmb24} approaches.

Despite this progress, two critical research gaps remain. First, most studies primarily evaluate algorithmic innovations without systematically considering the influence of dataset properties on performance \cite{SurveyofBigDataManagement16}. Real-world datasets often exhibit properties such as a small size, high domain specificity, missing values, and low overlap between sources, which differ from those of the relatively clean and standardized datasets that are commonly used in research. Neglecting these factors may result in methods that excel on benchmarks but underperform in practical enterprise settings being produced. Second, existing approaches have largely remained siloed: graph-based methods excel in capturing tabular structures, whereas knowledge-based methods use semantic context, yet few studies have attempted to combine these complementary strengths \cite{REMA20,bizer2012,Data:toShareornot24}.

Motivated by these limitations, this study aims to advance data integration methods that are robust across diverse dataset properties and are practically useful in enterprise workflows. Our contributions are twofold:

\begin{enumerate}
\item We propose a contextual graph embedding method that extends existing graph-based approaches by incorporating column- and token-level similarities and weighted random walks guided by contextual information.

\item We examine how dataset properties—specifically the domain specificity, data size, missing rate, and overlap rate—affect the performance of data integration tasks.
\end{enumerate}

These contributions are structured around the following research questions, which guide the design of our experiments and analysis:

\textbf{RQ1}: Which dataset properties influence SM and ER performance most significantly?

\textbf{RQ2}: Which types of algorithms are capable of maintaining robust and accurate matching performance across heterogeneous datasets?

\textbf{RQ3}: What are the boundary conditions under which matching performance can be consistently maintained, and where do existing approaches begin to fail?

\section{Related Studies and Our Approach}
\subsection{Rule-Based Approaches}
Rule-based methods are the most basic methods for handling data integration tasks. In these methods, experts establish logical conditions or custom rules to determine when two columns or records should be matched, and they deduce attribute mappings by assessing the column name similarity and data types, and employing statistical methods\cite{Cupid01}.\par 
Nevertheless, the success of heuristic matching rules is largely dependent on the quality of the manually created rules. Crafting and fine-tuning these rules are challenging and labor-intensive tasks require significant domain knowledge. To overcome this knowledge engineering challenge, recent studies have investigated semi-automated methods for creating or improving matching rules\cite{SynthesizingER17,Magellan16,GDD25}.\par 
In summary, although purely rule-based methods are becoming less prevalent in achieving state-of-the-art performance, they still provide a basis for interpretable matching and are frequently utilized in industry as preliminary matching filters or parts of larger workflows.\par
Compared with earlier probabilistic record linkage models that estimate match probabilities using handcrafted comparison vectors, our approach maintains interpretability while making use of graph-based structural cues that such models cannot easily capture. Similarly, although graph neural networks (GNNs) can learn rich multi-hop dependencies, they typically require large amounts of labeled data and extensive tuning [20]. In contrast, our embedding method operates in an unsupervised and lightweight manner, incorporating contextual cues from pre-trained language models and positioning itself as a middle ground between classical probabilistic linkage and fully supervised GNN-based matching.

\subsection{Knowledge-Based Approaches}
Knowledge-based approaches utilize external resources such as ontologies or dictionaries to aid in data integration tasks\cite{OntologyMatching04}. These techniques can detect synonyms using lexical knowledge bases such as WordNet\cite{Wordnet95}, or create connections through predefined mapping dictionaries. Recently, pretrained embeddings such as Word2Vec\cite{word2vec17} and GloVe\cite{Glove14} have been used to assess the semantic similarity of column names. Recent research has also explored the use of large pretrained language models for schema matching. For example, Li et al. fine-tuned Transformer models for entity matching by incorporating contextual knowledge from extensive text corpora\cite{DeepER20}.\par
Knowledge-based methods can significantly enhance the matching quality when appropriate resources are accessible, and have become a standard element in many contemporary data integration solutions. However, high-quality domain-specific knowledge is not always available and creating or selecting a suitable ontology or knowledge graph requires considerable effort. Even general resources such as Wikipedia or WordNet may have gaps or use terminology that does not align with the target datasets\cite{REMA20}. In addition, the table structure is often ignored in matching, resulting in performance that may vary significantly depending on the dataset characteristics.

\subsection{Graph-Based Approaches}
Graph-based methods have gained prominence with advancements in machine learning. These methods transform data into a graph format, learn embeddings on it, and then perform matching tasks. A notable example is EmbDI\cite{EmbDI20}, which creates a tripartite graph and executes data integration by learning node embeddings. REMA\cite{REMA20} incorporates attribute information as nodes using ARC graphs, and highlights node relationships through random walks to facilitate matching.\par
In addition to embedding techniques, other graph-based strategies address ER by directly examining the graphs of record linkages. For instance, Kirielle et al. addressed complex ER in contexts in which entities exhibit temporal changes and relationships\cite{Unsupervised23}. These methods excel by utilizing the inherent structures of table data, which are consistently available. However, they may not perform as well on datasets in which the data structure is challenging to discern, such as those with extensive numerical data or small datasets.

\subsection{Research Objective}
Previous research has predominantly focused on developing new algorithms, without thoroughly evaluating how dataset characteristics influence the effectiveness of SM and ER. For instance, Mudgal et al. introduced the use of deep learning for entity matching; however, their experiments were mainly conducted on standard benchmark datasets, rather than on noisy or domain-specific datasets \cite{mudgal2018deep}. More recent reviews \cite{christophides2020overview} have explicitly recognized that the variability of real-world data has not been well explored, and ensuring the robustness of methods across diverse dataset characteristics remains a significant challenge.\par 

Ignoring these dataset characteristics may result in methods being created that perform well on carefully curated benchmarks but fail in real-world scenarios, in which datasets are often small, domain-specific, partially overlapping, and incomplete. This discrepancy contributes to the limited use of automated data integration methods in business environments.\par

Our study addresses these gaps by proposing a hybrid framework that augments graph embeddings using contextual information from column descriptions and external knowledge. We further conduct systematic experiments across datasets with four properties that could significantly impact the integration performance: (i) the \textbf{data size}, as smaller datasets may not provide enough evidence for dependable matching; (ii) \textbf{domain specificity}, which can restrict the use of general-purpose lexical or ontology resources\cite{Principles12}; (iii) the \textbf{missing rate}, where a high percentage of missing values can introduce noise and diminish the reliability of attribute comparisons\cite{rahm2000data}; and (iv) the \textbf{overlap rate} between datasets, as a low overlap can decrease the effectiveness of schema- or instance-based similarity\cite{jimenez2011logmap}. Previous surveys on big data management have emphasized these properties frequently occur together in real-world situations, exacerbating the integration challenge\cite{SurveyofBigDataManagement16}.

\begin{figure*}[htbp]
    \centering
    \includegraphics[width=\linewidth]{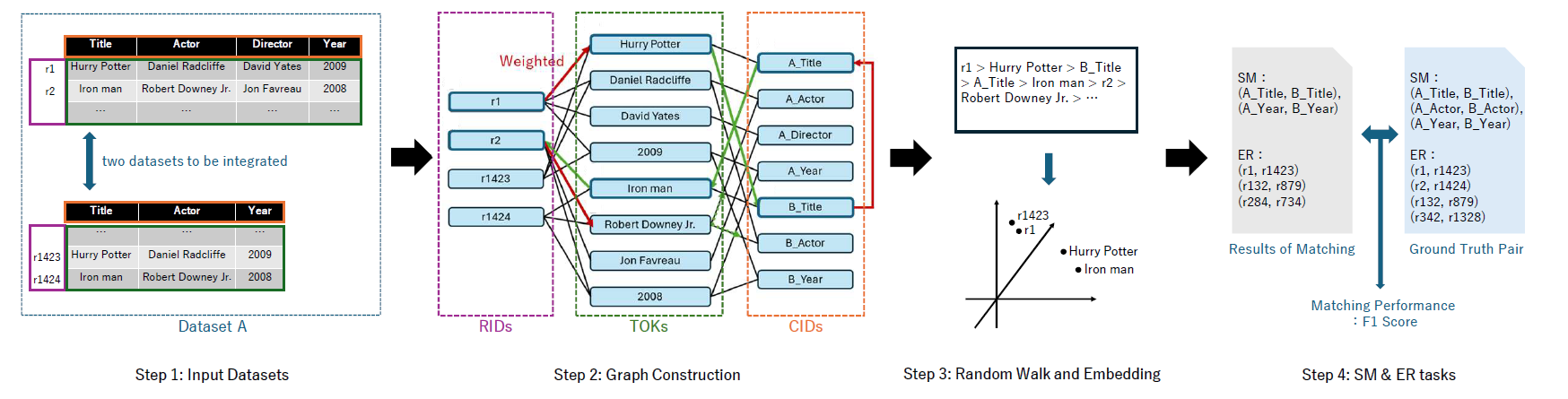}
    \caption{Architecture of the proposed contextual 4-partite graph embedding framework. Steps 1 \& 2: The model extends the tripartite graph—with RIDs, TOKs, and column CIDs—into a 4-partite graph by introducing weighted edges between the CIDs based on schema- and instance-level similarity. Step 3: The method learns enriched embeddings that capture both structural and contextual information through token merging and weighted random walks guided by column importance. Step 4: These embeddings are then used for SM and ER, improving robustness across heterogeneous datasets with varied dataset properties.}
    \label{Fig: Method Flow}
\end{figure*}

\section{Method}
Our method integrates graph- and knowledge-based approaches (Fig. \ref{Fig: Method Flow}). It specifically incorporates column descriptions and external knowledge into the graph structure. Following the procedure below, we learn embeddings for columns and rows, and then perform SM and ER using the cosine similarity. The architecture of the proposed contextual 4-partite graph embedding framework is illustrated in Fig. \ref{Fig: Method Flow}, and the complete graph construction process is detailed in Algorithm 1.

\subsection{Graph Construction and CID Similarity Measurement}
EmbDI is a notable graph-based approach that employs a tripartite graph to encapsulate the structure of datasets, thereby enhancing the data integration capabilities. This graph depicts the interconnections among entities within a dataset and comprises three distinct node types: token nodes (TOKs), record identifier nodes (RIDs), and column identifier nodes (CIDs). TOKs denote the values present in individual cells, RIDs serve to uniquely identify tuples and represent specific records, and CIDs are tasked with uniquely identifying each column (attribute) in the dataset. Despite its advantages, EmbDI is restricted to addressing only the structural elements of tabular data and does not account for contextual information regarding the dataset. Consequently, it may encounter difficulties when it is applied to datasets that are largely numerical or that contain numerous missing values.\par

Our proposed approach employs multiple datasets (two in our experiments) to generate a tripartite graph similar to EmbDI. To incorporate contextual information, EmbDI is extended by developing a 4-partite graph in which weighted edges are established between the CIDs based on the column similarity (Steps 1 and 2 in Fig. \ref{Fig: Method Flow}). Previous research has indicated that integrating schema- and instance-based information can result in high performance\cite{HybridEmbeddings23}. Therefore, this study considers both types of information. The method for assessing column similarity is detailed below.

\begin{figure}[!t]
\hrule \vspace{0.35em}
\textbf{Algorithm 1\; GenerateFourPartiteGraph}
\vspace{0.35em}

\begin{algorithmic}[1]
\STATE \textbf{Input:} two relational datasets $A_1,A_2$;\; threshold $\tau_{\text{cid}}$
\STATE \textbf{Output:} 4-partite graph $G$
\STATE let $G =$ empty graph
\FORALL{$attr$ in columns($A_1$) $\cup$ columns($A_2$)} 
  \STATE $G.\mathrm{addNode}(\mathrm{CID}_{attr})$
\ENDFOR
\FORALL{$T \in \{A_1,A_2\}$}
  \FORALL{$rid$ in rows($T$)}
    \STATE $G.\mathrm{addNode}(\mathrm{RID}_{rid})$
    \FORALL{cell $(rid,attr)$ with value $val$}
      \STATE For each cell $(rid,attr)$ with value $val$, tokenize $val$; for each resulting token $tok$, add a node $\mathrm{TOK}_{tok}$ and edges  $\mathrm{RID}_{rid}\!\leftrightarrow\!\mathrm{TOK}_{tok}\!\leftrightarrow\!\mathrm{CID}_{attr}$.
    \ENDFOR
  \ENDFOR
\ENDFOR
\FORALL{$i$ in columns($A_1$)} \FORALL{$j$ in columns($A_2$)}
  \STATE $S_{ij} \leftarrow \mathrm{Similarity}(i,j)$ 
  \IF{$S_{ij} \ge \tau_{\text{cid}}$}
    \STATE $G.\mathrm{addEdge}(\mathrm{CID}_i,\mathrm{CID}_j,\text{weight}=S_{ij})$
  \ENDIF
\ENDFOR \ENDFOR
\STATE \textbf{Output:} $G$
\end{algorithmic}

\vspace{0.35em}\hrule
\label{alg:fourpartite}
\end{figure}

\noindent\textbf{}
\noindent\textbf{}
\vskip0.5\baselineskip
\noindent\textbf{Schema-Based Similarity}\par
Schema-based similarity utilizes information such as the names of columns and their descriptions. For instance, consider the similarity between the column names $e_i$ in dataset $A_1$ and $e_j$ in dataset $A_2$. Furthermore, assume that metadata provide column descriptions labeled $\mathrm{desc}_i$ and $\mathrm{desc}_j$ for $e_i$ and $e_j$, respectively. To capture semantic connections beyond basic string matching, we combine each column name with its description and feed the resulting sentence into Sentence-BERT (sentence-transformers/paraphrase-multilingual-mpnet-base-v2)\cite{SentenceBERT19}. This model, which has been pretrained on semantic textual similarity and paraphrase detection tasks, produces dense embeddings $s_i$ and $s_j$ that capture the semantic similarity between columns.\par
Consider a scenario in which $e_i$ represents ``funding" and $\mathrm{desc}_i$ is described as ``the amount of financial support received." The combined input would be ``funding which means the amount of financial support received." This approach ensures that both the column label and its contextual meaning are incorporated into the embedding. The similarity $S_{\mathrm{schema}}$ between $e_i$ and $e_j$ is defined as the average of two complementary metrics: (1) the normalized cosine similarity between the embeddings $s_i$ and $s_j$ ($s_{\mathrm{cos}}$), which measures the semantic similarity in vector space, and (2) the normalized Levenshtein similarity\cite{lcvenshtcin1966binary} between the column names ($s_{\mathrm{lev}}$), which assesses the surface-level string similarity (Eqs. (\ref{s_cos})--(\ref{s_schema})). By integrating these metrics, the method considers both the semantic proximity and lexical overlap, thereby enhancing the robustness when column names are brief or unclear.
\begin{equation}
\label{s_cos}
    s_{\mathrm{cos}} = \cfrac{\cos(s_i, s_j) + 1}{2} \in [0,1]
\end{equation}

\begin{equation}
\label{s_lev}
    s_{\mathrm{lev}} = 1 - \cfrac{\operatorname{lev}(e_i, e_j)} {\max\{\lvert e_i\rvert, \lvert e_j\rvert\}} \in [0,1]
\end{equation}

\begin{equation}
\label{s_schema}
    S_{\mathrm{schema}}  = \operatorname{avg}\!\big( s_{\mathrm{cos}},\,  s_{\mathrm{lev}}  \big) \in [0,1]
\end{equation}

\vskip0.5\baselineskip
\noindent\textbf{Instance-Based Similarity}\par

Instance-based similarity aims to encapsulate the traits of cell values within a column. To achieve this, we construct $\mathbf{v}_i$ and $\mathbf{v}_j$, which represent the instance-based embeddings for columns $e_i$ and $e_j$, respectively.\par

First, the instance features are categorized into numeric and character features. For numeric columns, a feature vector $\mathbf{v}_i^{(\mathrm{num})}$ is used to summarize the distribution of valid values (Eq.~(\ref{v_num})). This vector includes statistics such as the mean, minimum, maximum, variance, and standard deviation. These statistics describe the overall distribution shape and capture the numerical similarities across columns. Such statistical features are valuable because they provide a concise representation of the distribution, enabling column comparisons even when the raw values differ in scale or units.
\begin{equation}
\label{v_num}
    \mathbf{v}_i^{(\mathrm{num})}
  = \big[\, \bar{x},\ \min(x),\ \max(x),\ \mathrm{Var}(x),\ \mathrm{SD}(x)\ \big]
\end{equation}

In contrast, character features focus on surface-level patterns. The feature vector $\mathbf{v}_i^{(\mathrm{char})}$ includes the proportions of spaces, punctuation, special symbols, and numbers, averaged across the column (Eq.~(\ref{v_char})). The four terms $r_{\mathrm{space}}$, $r_{\mathrm{punc}}$, $r_{\mathrm{special}}$, and $r_{\mathrm{digit}}$ represent the per-cell \textit{ratios} of these character categories (whitespace, punctuation, non-alphanumeric symbols, and digits); each \textit{r} is calculated for each cell as ``count of the class / total characters in the cell," and then averaged over the column to derive a column-level statistic. These features are used to identify structural or formatting patterns in values, such as identifiers, codes, or text fields.
\begin{equation}
\label{v_char}
    \mathbf{v}_i^{(\mathrm{char})}
  = \big[\, r_{\mathrm{space}},\ r_{\mathrm{punc}},\ r_{\mathrm{special}},\ r_{\mathrm{digit}} \,\big]
\end{equation}\par

Because these features depend on the type, any features that do not apply to a specific column type are represented by a fixed-length vector filled with $-1$, indicating that the feature is not valid. The final instance embedding $\mathbf{v}_i$ is created by concatenating the numeric and character feature vectors, as follows:
\begin{equation}
\label{v_i}
    \mathbf{v}_i
  = \big[
      \mathbf{v}_i^{(\mathrm{num})}
      \mathbf{v}_i^{(\mathrm{char})}
    \big].
\end{equation}\par

The instance-based similarity $S_{\mathrm{instance}}$ between two columns is calculated as the mean of two components: (i) the cosine similarity $v_{\mathrm{cos}}$ of their feature vectors, and (ii) the relative difference measure $s_{\mathrm{diff}}$:
\begin{equation}
\label{s_instance}
    S_{\mathrm{instance}}
  = \tfrac{1}{2}\, (v_{\mathrm{cos}} + s_{\mathrm{diff}}),
\end{equation}

where $v_{\mathrm{cos}}$ is the normalized cosine similarity:

\begin{equation}
    v_{\mathrm{cos}}
  = \cfrac{\cos(\mathbf{v}_i,\mathbf{v}_j) + 1}{2}
  \in [0,1],
\end{equation}

\noindent and $s_{\mathrm{diff}}$ is a similarity score based on the relative differences between vector elements:

\begin{equation}
\label{delta_k}
    \delta_k
  = \cfrac{\lvert v_{ik} - v_{jk} \rvert^{\beta}}{\,v_{ik} + v_{jk} + \varepsilon\,},
\qquad
d = \operatorname{mean}_{k \in \mathcal{K}}(\delta_k)
\end{equation}

\begin{equation}
\label{s_diff}
    s_{\mathrm{diff}}
  = 1 - \min\{1, d\}
  \in [0,1].
\end{equation}\par

In Eq.~(\ref{delta_k}), $v_{ik}$ represents the $k$-th element of the column feature vector $\mathbf{v}_i$ and $\mathcal{K}$ is the set of indices for the relevant features, excluding entries marked as inapplicable with $-1$. The operator $\operatorname{mean}_{k\in \mathcal{K}}(\cdot)$ calculates the arithmetic mean over the indices $k\in \mathcal{K}$ and $d$ denotes the average relative gap. Consequently, $s_{\mathrm{diff}}$ translates the average relative gap $d$ into a similarity score within the range [0,1] via clipping and inversion. The exponent $\beta<1$ reduces the impact of significant outliers, and features deemed inapplicable (padded with $-1$) are excluded from the calculation. Although the cosine similarity measures the directional closeness of feature vectors, it does not account for variations in magnitude. The relative-difference term $s_{\mathrm{diff}}$ complements the cosine similarity by considering the differences in scale and magnitude between columns. This enhances the robustness of the measure when datasets vary in units, ranges, or levels of missing data. In addition, because $s_{\mathrm{diff}}$ is based on relative rather than absolute gaps, it remains scale invariant, making it suitable for integration tasks involving heterogeneous datasets.

\vskip0.5\baselineskip
\noindent\textbf{Similarity Merger}\par
The total similarity $S_{ij}$ between columns $e_i$ and $e_j$ is characterized as a weighted blend of the schema-based similarity $S_{\mathrm{schema}}$ (Eq.~(\ref{s_schema})) and instance-based similarity $S_{\mathrm{instance}}$ (Eq.~(\ref{s_instance})), as follows:
\begin{equation}
\label{eq_s}
    S_{ij} \;=\; \alpha\, S_{\mathrm{schema}} \;+\; (1-\alpha)\, S_{\mathrm{instance}} \quad (\,0 \le \alpha \le 1\,),
\end{equation}

\noindent where the weight $\alpha$ is constrained to lie between $\alpha_{\min}$ and $\alpha_{\max}$ and is computed as follows:
\begin{equation}
\label{alpha}
    \alpha \;=\; \alpha_{\max} \;-\; (\alpha_{\max}-\alpha_{\min})\, r_{\mathrm{value}}
.
\end{equation}
In this context, $r_{\mathrm{value}}$ denotes the confidence level associated with instance-based similarity. A higher $r_{\mathrm{value}}$ suggests that the features of the instance are more dependable. For instance, when the dataset contains more records and fewer missing values, this results in a reduced $\alpha$ value. This implies that, when the evidence at the instance level is robust, the method assigns greater importance to $S_{\mathrm{instance}}$. However, if the instance-level evidence is weak, such as in the case of sparse or noisy data, the method leans more towards schema-level information.\par
We add a CID--CID edge only when the final similarity $S_{ij}$ exceeds a threshold $\tau_{\mathrm{cid}}$.

\subsection{Merging TOKs}
During graph construction, each cell value is treated as a node. In the EmbDI algorithm\cite{EmbDI20}, if a cell value appears in multiple tuples, the algorithm reuses the existing node rather than generating a new node. Nonetheless, in real-world scenarios, values may represent the same entity but manifest in slightly varied forms. For instance, two movie datasets may list actors’ names differently, such as ``Daniel Jacob Radcliffe" and ``Daniel Radcliffe." In these situations, combining similar TOKs within the graph can enhance the matching efficiency and reduce computational expenses. This merging step is crucial for real-world datasets in which differences in spelling, abbreviations, or formatting are prevalent and could otherwise lead to a fragmented graph representation.

The proposed method utilizes FastText (cc.en.300. vec)\footnote{\url{https://fasttext.cc/docs/en/crawl-vectors.html}} to assess the distance between TOKs. FastText was selected because, unlike conventional word embeddings such as Word2Vec\cite{word2vec17}, it encodes words as subword \textit{n}-grams. This feature enables the recognition of similarities between tokens with slight variations in spelling or morphology (such as plural forms, abbreviations, or misspellings), making it particularly effective in handling noisy or diverse tabular data. By embedding tokens into this semantic space and calculating their cosine similarity, this method can detect and combine nodes that are likely to represent the same entity, thereby minimizing redundancy and improving the graph structure quality.

\subsection{Random Walks and Learning of Embedding Representations}
We perform random walks on the 4-partite graph to generate embedding representations from the resulting sequences, which are then used for SM. These random walks involve creating sequences of nodes by starting at each node in the 4-partite graph and moving to neighboring nodes based on the transition probabilities. Each node is selected as a starting point with a predetermined probability. The process of randomly transitioning to adjacent nodes continues until a specific sequence length is reached, thereby producing multiple sequences. In contrast to EmbDI, our method applies weights to certain transitions during the random walks. These weighted random walks occur in two types of transitions: from RIDs to TOKs and from CIDs to their neighbors.

\vskip0.5\baselineskip
\noindent\textbf{Transitions from RIDs to TOKs} \par
The significance of the columns or cells can differ significantly when interpreting the content of a dataset. For instance, in a movie dataset, the ``Title" column typically offers more valuable information than the ``Year" column. Consequently, conducting random walks more frequently through key columns or cells can integrate more pertinent information into the sequences, potentially enhancing the matching performance. Our method initially assigns weights to the CIDs. By applying weights ($c$) and aggregating the four indicators signifying the importance of the columns (missing rate, data type, data variety, and column generality), the weight of node $i$, represented as $w_i$, is determined as follows:
\begin{equation}
\label{w_a}
w_i = c_m s_{\mathrm{miss}} + c_l s_{\mathrm{ling}} + c_v s_{\mathrm{variety}} + c_f s_{\mathrm{freq}}.
\end{equation}
\noindent In this context, $s_{\mathrm{miss}}$ changes based on the percentage of missing cells within a column. Columns deemed more significant typically exhibit lower missing data rates; thus, $s_{\mathrm{miss}}$ increases as the missing rate decreases. The score $s_{\mathrm{ling}}$ indicates whether the cells contain linguistic or numerical information. Graph-based methods, including ours, often underperform on datasets dominated by numerical data. Hence, $s_{\mathrm{ling}}$ increases with a higher proportion of linguistic data. The score $s_{\mathrm{variety}}$ measures the diversity of the cell values in a column. Columns of greater importance are anticipated to have a wider range of distinct values, leading to a higher $s_{\mathrm{variety}}$. Finally, $s_{\mathrm{freq}}$ assesses the generality of a column name. Columns with more generic names are likely to be more significant because they frequently appear in various datasets. To assess the generality, we utilize the Meta Kaggle dataset\footnote{\url{https://www.kaggle.com/datasets/kaggle/meta-kaggle}}, which provides metadata on datasets that are shared on Kaggle. If a column name ranks among the top 100 most common variables in metadata in the Meta Kaggle dataset, we determine $s_{\mathrm{freq}}$ based on its frequency. The weights $c_m$, $c_l$, $c_v$, and $c_f$ are assigned to the four respective scores.\par

Subsequently, we determine the weight of each TOK by calculating the weighted average of the weights of its adjacent CIDs. This TOK weight guides the random walk during transitions from RIDs to TOKs. For instance, in Step 2 of Fig. \ref{Fig: Method Flow}, when moving from the node ``r1," among the four possible destination nodes, the probability of transitioning to the token ``Harry Potter" is determined by the following equation:
\begin{equation}
P_{\mathrm{HarryPotter}} = \tfrac{1}{2}\, (w_{\mathrm{A\_Title}} + w_{\mathrm{B\_Title}}).
\end{equation}

\vskip0.5\baselineskip
\noindent\textbf{Transitions from CIDs}\par
Transitions from CIDs involve not only adjacent TOKs but also CIDs linked by similarity edges. For instance, in Step 2 of Fig. \ref{Fig: Method Flow}, when moving from the node ``B\_Title," the transition target is selected from three possibilities: the TOKs ``Harry Potter" and ``Iron Man," and the CID ``A\_Title," which is linked by an edge. The random walk transition probability is influenced by the weight assigned to the edge leading to ``A\_Title." Conversely, all edges connecting the TOKs have the same weight.

\vskip0.5\baselineskip
Finally, we derive embedding representations from the sequences generated by these weighted random walks. We consider these sequences as a corpus of documents and utilize the Skip-Gram model from Word2Vec to generate embedding vectors for each node. The embeddings have 300 dimensions and a context window of 3. By integrating weights into the random walks, the resulting embeddings not only reflect the structural relationships but also highlight the relative significance of various columns and tokens, thereby producing more insightful representations for SM.\par

\begin{table*}[htbp]
\caption{Dataset properties used in Experiment 1.}
\centering
\renewcommand{\arraystretch}{1.2}
\begin{tabular}{lcrrrr}
\toprule
Dataset & Domain specificity & \# data size  & \% missing & \% overlap (column) & \% overlap (row) \\
\midrule
Dataset A    &  low & 52,545  & 5.84 & 62.40 & 7.12 \\
Dataset B    &  low & 103,243 & 9.43 & 38.10 & 4.35 \\
Dataset C    &  mid & 104,235 & 10.60 & 64.22 & 6.43 \\
Dataset D    &  high & 53,942 & 15.38 & 44.53 & 5.23 \\
Dataset E    &  high & 42,353 & 19.64 & 68.55 & 20.43 \\
\bottomrule
\end{tabular}
\label{table:dataset-properties}
\end{table*}

\section{Experiments}
\subsection{Experimental Methods and Evaluation}
This study evaluated the effectiveness of the proposed contextual graph embedding techniques for two key tasks of data integration: SM and ER. Whereas earlier research often focused solely on algorithmic choices as the primary factor affecting performance, this study considered two elements: (1) variations in the algorithms employed and (2) the specific characteristics of the target datasets. We conducted two experiments to examine the effects of these elements.

\subsection{Datasets}
Experiment 1 evaluated the impact of algorithmic choices on performance. We prepared five datasets (A--E) with distinct properties, performed the SM and ER tasks, and verified their performances. The summary statistics for datasets A--E are presented in Table~\ref{table:dataset-properties}.\par 
Experiment 2 isolated the impact of the dataset properties. For each of the four properties, we generated controlled variants that varied this property stepwise while maintaining the remaining three as similar as possible. We then ran the SM and ER tasks on these variants to quantify the effects of each property.\par 
Datasets A--E were each constructed by editing different source datasets (public or anonymized enterprise) so that every dataset in A--E originated from a distinct base. Similarly, datasets F--H were derived by editing different base datasets to represent varying levels of domain specificity.\par
In contrast, the dataset triples used in Experiment 2 (I--K, L--N, and O--Q) were each generated from a single common base dataset per triple. Within each triple, we varied one property stepwise while maintaining the remaining properties as similar as possible through controlled editing.

\subsection{Experiment 1: Effects of Algorithm Differences on Performance}
To evaluate the effects of algorithmic differences on the data integration performance, we compared three representative approaches for SM and ER for each dataset pair.

\vskip0.5\baselineskip
\noindent \textbf{Proposed Method (Ours)}: A graph-based approach inspired by EmbDI\cite{EmbDI20}, incorporating the following three innovations: (1) Weighted CID edges that incorporate schema- and instance-level similarity between columns., (2) TOK merging using semantic similarity (FastText) to unify variant expressions of the same entity., (3) Weighted random walks guided by column importance, enabling embeddings that reflect both structural and contextual information.

\vskip0.5\baselineskip
\noindent \textbf{EmbDI} (Baseline Graph-Based Method): The original graph embedding approach \cite{EmbDI20}, which constructs a tripartite graph of records, tokens, and columns, and learns embeddings through uniform random walks. Although effective in capturing structural relationships in tabular data, EmbDI does not incorporate contextual information or external knowledge, thereby limiting its robustness under noisy or domain-specific conditions.

\vskip0.5\baselineskip
\noindent \textbf{GPT-5} (Large Language Model (LLM) Baseline): A state-of-the-art LLM prompted directly to perform SM and ER by identifying column correspondences between dataset pairs with the prompt below. Although it can leverage broad linguistic knowledge, GPT-5 lacks intrinsic mechanisms for capturing tabular structures or statistical patterns, making it an important baseline for testing the limits of LLMs in data integration tasks.

\begin{figure}[h]
    \centering
    \begin{tcolorbox}[
    enhanced,
    colframe=black, 
    colback=white, 
    coltitle=black, 
    colbacktitle=white, 
    boxed title style={frame hidden}, 
    fonttitle=\bfseries,
    left=1mm,    
    right=1mm,   
    top=1mm,     
    bottom=0mm,  
    boxsep=1mm,  
    before skip=1mm, 
    after skip=1mm   
    ]
    \small{\small{
    \texttt{Perform schema matching on the following two datasets. Schema matching is the task of finding corresponding relationships between columns in two different datasets to be merged. For example, it involves determining that the columns ``Customer ID" in dataset A and ``Client Number" in dataset B refer to the same concept. List all applicable pairs as shown in the example below. Column ``temperature" of dataset A and column ``temp" of dataset B.}}}
    \end{tcolorbox}
\end{figure}

Subsequently, we determined the F1 score by comparing the predicted matches with the actual pairs and utilized this as the performance measure for the SM and ER (Step 4 of Fig. \ref{Fig: Method Flow}).

\subsection{Experiment 2: Effects of Dataset Properties on Performance}
Previous research has not extensively explored which attributes of datasets affect the matching performance or the degree of their influence. In Experiment 2, we assessed how dataset characteristics affect thee data integration performance by comparing EmbDI with the proposed approach using datasets with varying properties. We focused on four properties as evaluation criteria: the domain specificity, data size, missing rate, and overlap rate.

\vskip0.5\baselineskip
\noindent \textbf{Domain Specificity}\par
Domain specificity pertains to the distinctiveness of the columns and content within a dataset. For instance, a dataset featuring highly specialized terms that are unique to a particular organization is considered to have high domain specificity, whereas a dataset primarily composed of generic columns (such as ID, Name, Address, or Gender) is considered to have low domain specificity. We used internal company datasets as examples of high-domain-specificity data, Kaggle datasets as examples of mid-domain-specificity data, and datasets from previous research as examples of low-domain-specificity data. We then generated additional datasets by modifying these sources.

\vskip0.5\baselineskip
\noindent \textbf{Data Size}\par
The data size is defined as the number of columns multiplied by the number of rows (total cells).

\vskip0.5\baselineskip
\noindent \textbf{Missing Rate}\par
The missing rate is defined as the proportion of unfilled cells within a dataset. Empty cells may indicate either unavailable information or data that were never collected. Because it is difficult to automatically distinguish between these cases, we generally treat missing cells as unavailable data and remove those that clearly represent non-applicable cases.

Second, multiple methods are available for representing missing values. In addition to leaving the cells empty, placeholders such as NA, null, or 0 can be used. In this study, we transformed all missing values into empty cells. However, in some datasets, blank spaces act as indicators. For instance, in a column labeled ``Foreign Nationality," rows with a match might be denoted by ``1," whereas unmatched rows remain blank. In this context, the blank indicates ``no match" rather than missing data. For such datasets, we replaced blanks with ``0."

Third, the distribution of missing cells can affect how datasets are understood and their effectiveness in matching tasks. In real-world datasets, missing values tend to cluster in specific rows or columns rather than being distributed randomly. Although this study did not explore this issue in depth, our dataset was designed to mirror realistic patterns of missing values, with these values mainly appearing in less significant columns. In contrast, columns that are essential for identifying entities, such as ID or Name, typically showed lower instances of missing values.

\vskip0.5\baselineskip
\noindent \textbf{Overlap Rate}\par
The overlap rate quantifies the extent of the shared content between two datasets to be combined. It is calculated separately for the columns and rows, as follows:
\begin{equation}
\label{overlap_column}
\mathrm{overlap}_{\mathrm{column}} = \frac{t_{\mathrm{SM}}}{n_{\mathrm{column}}},
\end{equation}
\begin{equation}
\label{overlap_row}
\mathrm{overlap}_{\mathrm{row}} = \frac{t_{\mathrm{ER}}}{n_{\mathrm{row}}}.
\end{equation}
In Eqs. (\ref{overlap_column}) and (\ref{overlap_row}), $t_{\mathrm{SM}}$ denotes the number of ground-truth column pairs and $t_{\mathrm{ER}}$ denotes the number of ground-truth record pairs.
$n_{\mathrm{column}}$ is the number of columns in the smaller of the two datasets, and $n_{\mathrm{row}}$ is the number of rows in the smaller of the two datasets.

\begin{table*}[htbp]
\caption{Results of Experiment 1: SM and ER performance (F1).\\ (The best scores are highlighted in bold.)}

\centering
\setlength{\tabcolsep}{6pt}
\renewcommand{\arraystretch}{1.2}
\begin{tabular}{|l|ccc|ccc|}
\hline
\multicolumn{1}{|c|}{\textbf{Dataset}} &
  \multicolumn{3}{c|}{\textbf{SM (F1)}} &
  \multicolumn{3}{c|}{\textbf{ER (F1)}} \\
\cline{2-7}
\multicolumn{1}{|c|}{} &
  \textbf{Ours} & \textbf{EmbDI} & \textbf{GPT-5} &
  \textbf{Ours} & \textbf{EmbDI} & \textbf{GPT-5} \\
\hline
Dataset A & \textbf{0.84} & 0.74 & 0.64 & \textbf{0.84} & 0.80 & 0.39 \\
Dataset B & \textbf{0.84} & \textbf{0.84} & 0.71 & \textbf{0.82} & 0.78 & 0.52 \\
Dataset C & \textbf{0.82} & \textbf{0.82} & 0.75 & \textbf{0.84} & 0.77 & 0.43 \\
Dataset D & \textbf{0.80} & 0.66 & 0.68 & \textbf{0.79} & 0.69 & 0.35 \\
Dataset E & \textbf{0.71} & 0.60 & 0.39 & \textbf{0.76} & 0.60 & 0.25 \\
\hline
\end{tabular}
\label{tab:sm-er-results}
\end{table*}

\section{Results and Discussion}  

\subsection{Results of Experiment 1: Addressing RQ1}
As outlined in Table~\ref{table:dataset-properties}, the experimental datasets varied in terms of the domain specificity, size, missing data, and overlap. The SM/ER results are listed in Table~\ref{tab:sm-er-results}. Our proposed approach performed on par with or better than EmbDI across most datasets, and consistently outperformed GPT-5 in both the SM and ER. This indicates that integrating column descriptions and external knowledge enhances the robustness beyond structural-only graph embeddings, whereas LLMs alone are inadequate for dependable SM/ER. Performance improvements were particularly significant for datasets with a high percentage of numerical values or substantial intra-column variation. In these scenarios, EmbDI was constrained by its dependence on exact token matches, whereas the proposed method used contextual embeddings to identify latent similarities.\par
However, the proposed method introduced some errors into columns with lexically similar but semantically different labels (e.g., \textit{Max\_Temp} vs. \textit{Min\_Temp}). In these instances, the contextual similarity inflated the matching scores, leading to false positives. This suggests that fine-tuning Sentence-BERT for domain-specific matching data is necessary to reduce such errors. In addition, both methods struggled with datasets characterized by high missing rates, insufficient metadata, or strong domain-specific terminology. These findings underscore the inherent limitations of fully automated matching and imply that human review remains essential in enterprise environments.\par
Finally, GPT-5 underperformed compared with both graph-based methods. This likely reflects the challenge of capturing the structural and statistical patterns of tabular data using general-purpose LLMs. Although recent studies have explored the conversion of tables into textual forms for LLM processing~\cite{Rematch24}, our findings suggest that dedicated algorithms are still required.

\subsection{Results of Experiment 2: Addressing RQ2 and RQ3}  
\begin{figure}[htbp]
    \centering
    \includegraphics[width=\linewidth]{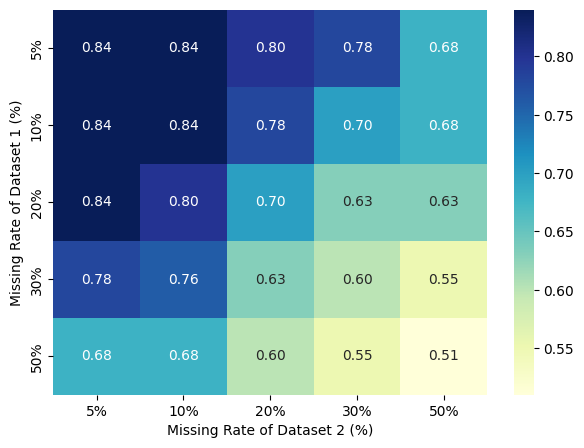}
    \caption{SM performance under varying missing rates.}
    \label{Fig: result_SM}
\end{figure}

\begin{figure}[htbp]
    \centering
    \includegraphics[width=\linewidth]{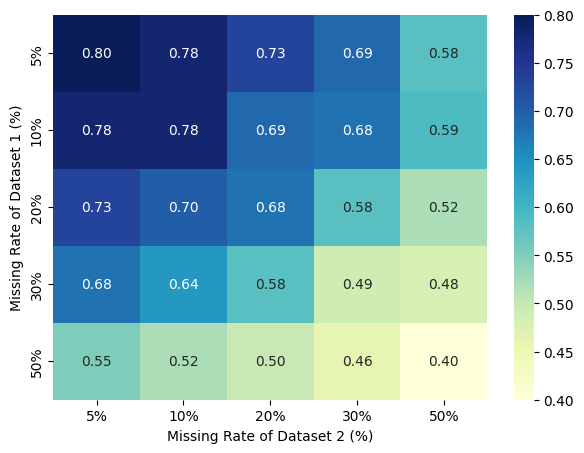}
    \caption{ER performance under varying missing rates.}
    \label{Fig: result_ER}
\end{figure}

Experiment 2 investigated how dataset properties affect SM/ER performance (RQ2) and compared the behavior of the proposed method and EmbDI across various property conditions (RQ3). In Experiment 1, the performance of GPT-5 was significantly lower than that of both the proposed method and EmbDI. This finding led Experiment 2 to focus solely on these two graph-based methods, facilitating a more direct comparison of the various dataset properties. Table~\ref{tab:datasets} summarizes the datasets used, which varied in terms of the domain specificity, data size, overlap rate, and missing rate. Figs.~\ref{Fig: result_SM} and~\ref{Fig: result_ER} depict the performance trends when the missing rates were systematically adjusted.

\begin{table*}[t]
\caption{Results of Experiment 2: Performance of our method and EmbDI across datasets with different properties.\\ (The best scores are highlighted in bold.)}

\centering
\small
\renewcommand{\arraystretch}{1.2}
\begin{tabular}{|l|l|l|c|c|c|c|}
\hline
\multirow{2}{*}{\textbf{Property}} & 
\multirow{2}{*}{\textbf{Dataset}} &
\multirow{2}{*}{\textbf{Details}} &
\multicolumn{2}{c|}{\textbf{SM (F1)}} &
\multicolumn{2}{c|}{\textbf{ER (F1)}} \\ \cline{4-7}
& & & \textbf{Ours} & \textbf{EmbDI} & \textbf{Ours} & \textbf{EmbDI} \\ \hline

\multirow{3}{*}{Domain specificity} 
& Dataset F & Based on prior research dataset
& \textbf{0.85} & 0.69 & \textbf{0.84} & 0.70\\ \cline{2-7}
& Dataset G & Based on Kaggle datasets
& \textbf{0.80} & 0.67 & \textbf{0.76} & 0.63 \\ \cline{2-7}
& Dataset H & Corporate transaction data
& \textbf{0.67} & 0.57 & \textbf{0.75} & 0.55 \\ \hline

\multirow{3}{*}{Data size}
& Dataset I &  Data size = 300,000
& \textbf{0.78} & 0.72 & \textbf{0.80} & 0.68\\ \cline{2-7}
& Dataset J & Data size = 50,000
& \textbf{0.78} & 0.72 & \textbf{0.78} & 0.70 \\ \cline{2-7}
& Dataset K &  Data size = 5,000
& \textbf{0.75} & 0.65 & \textbf{0.76} & 0.69\\ \hline

\multirow{3}{*}{Overlap (column)} 
& Dataset L & Overlap = 10\%
& \textbf{0.74} & 0.69 & \textbf{0.38} & 0.20\\ \cline{2-7}
& Dataset M & Overlap = 30\%
& \textbf{0.68} & 0.60 & \textbf{0.70} & 0.68 \\ \cline{2-7}
& Dataset N & Overlap = 50\%
& \textbf{0.84} & 0.80 & \textbf{0.88} & \textbf{0.88} \\ \hline

\multirow{3}{*}{Overlap (row)}
& Dataset O & Overlap = 10\%
& \textbf{0.70} & 0.68 & \textbf{0.72} & 0.67 \\ \cline{2-7}
& Dataset P & Overlap = 30\%
& \textbf{0.71} & 0.68 & \textbf{0.78} & 0.74 \\ \cline{2-7}
& Dataset Q & Overlap = 50\%
& \textbf{0.74} & \textbf{0.74} & \textbf{0.70} & 0.64 \\ \hline

\end{tabular}
\label{tab:datasets}
\end{table*}

\par

To address RQ2, which asks which algorithms can maintain robust and accurate matching across diverse datasets, Experiment 2 highlighted the systematic impact of the dataset properties. As the specificity of the domain increased, the performance degraded, indicating that general tools such as Sentence-BERT embeddings are less effective for specialized vocabulary. There was a threshold effect concerning the data size: very small datasets hindered performance; however, beyond a certain size, larger datasets provided minimal additional advantages. The overlap rate significantly influenced the performance, with a low column overlap notably diminishing the results.\par
The rate of missing data was also crucial: the performance gradually declined up to approximately 20\% missingness, but a steep drop occurred between 30\% and 50\%. Even if one dataset had a low missing rate, a high rate in the other dataset reduced the performance of both the SM and ER because the effective overlap of informative evidence decreased. Beyond this rate, the missingness pattern was also important. As discussed in Section IV-D, real-world datasets show structured missingness (e.g., column- or row-wise clusters), and the performance depends on the affected attributes. When missing cells are concentrated in low-importance columns, the ER tends to degrade less; conversely, missingness in high-importance entity-identifying columns leads to significant drops. The ER was more affected by missingness than the SM, likely because the alignment records relied on multiple attribute values.\par
RQ3 explores the factors that guarantee consistent matching performance and pinpoints where current methods begin to falter. Compared with EmbDI, our method excelled in scenarios characterized by high numerical content, considerable intra-column diversity, or moderate overlap. In such cases, token-based random walks tend to lose contextual characteristics, as identical values rarely co-occur, resulting in sparse connections. The contextual embeddings mitigate this limitation by incorporating column descriptions and distributional statistics, thereby providing semantic and statistical cues even when token overlap is minimal. Conversely, when analyzing the matching outcomes, in some instances, EmbDI accurately recognized cases in which the column labels were lexically similar but semantically different, whereas the proposed method failed to do so. Importantly, in situations with significant data missingness or extreme domain specificity, both methods faced challenges, highlighting the necessity of semi-automated processes that combine machine matching with human validation.

\subsection{Summary Discussion}  
In summary, these results offer definitive answers to the three research questions:  
\vskip0.5\baselineskip
\noindent \textbf{RQ1}: The proposed approach either matches or surpasses EmbDI and significantly outperforms GPT-5 in SM and ER across diverse datasets. This demonstrates that adding contextual information to graph embeddings enhances robustness compared with using only graph- or LLM-based baselines.  
\vskip0.5\baselineskip
\noindent \textbf{RQ2}: The characteristics of the dataset—such as the size, domain specificity, overlap rate, and missing rate—have a significant impact on the matching performance. Notably, the performance declines significantly when the overlap is minimal or missingness is high, and the domain specificity restricts the effectiveness of general external resources.  
\vskip0.5\baselineskip
\noindent \textbf{RQ3}: The proposed method exhibits the most significant improvements in heterogeneous and noisy environments (e.g., high numerical content and diverse values), although EmbDI may outperform it in specific scenarios in which column labels are lexically similar but semantically different. These findings suggest that contextual embeddings offer higher recall at the risk of false positives, indicating the need for fine-tuned models or human-in-the-loop adjustments.  
\vskip0.5\baselineskip
\par
Overall, these findings underscore both the potential and limitations of contextual graph embeddings. Although they greatly enhance robustness in many real-world situations, careful management of ambiguous attributes and high-missingness data is crucial for effective implementation.

\section{Conclusion and Next Steps}  
This study explored the effects of dataset characteristics on data integration tasks and introduced a new embedding-based approach that integrates contextual information within a graph-based framework. We addressed three research questions by assessing SM and ER across datasets with varying properties. The results indicate that assessing algorithms without considering dataset properties can lead to an overestimation of their practical applicability. Our findings also suggest that effective methods should integrate multiple perspectives (structural and contextual) to achieve a balance between precision and recall.

In future research, we intend to (i) enhance the similarity metrics and their parameters to better differentiate semantically different but lexically similar columns, (ii) investigate additional dataset properties that may affect performance, and (iii) assess semi-automated workflows in which human experts verify and amend machine-generated matches. Furthermore, a more comprehensive analysis is required to understand the types of errors generated by each method; specifically, which column pairs are incorrectly matched and which true correspondences are often overlooked. This analysis will offer insights into the complementary strengths and weaknesses of graph-based, knowledge-based, and LLM approaches that inform the design of hybrid systems.

For semi-automated workflows, we foresee tasks in which human experts systematically review machine-generated matching outcomes. For instance, experts cross-check the suggested column pairs while directly examining the underlying datasets, thereby confirming the accuracy of the proposed matches and identifying missed pairs. This interactive verification process can ensure high reliability in practical applications and provide valuable feedback for the further enhancement of automated methods. We expect that these initiatives will aid in the development of data integration systems that are robust under real-world conditions.

\section*{Acknowledgment}
This study was supported by the joint research project with Infomart Corporation and JST PRESTO (JPMJPR2369).

\bibliographystyle{IEEEtran}
\bibliography{bigdata2025_arXiv}

\end{document}